\documentclass[12pt,a4paper]{article}
\usepackage{amsmath}
\usepackage{bm}
\usepackage{amsfonts}
\usepackage{graphicx}
\usepackage[normal]{caption2}
\usepackage{subfigure}
\usepackage{rotating}
\usepackage{citesort}
\usepackage{lscape}
\usepackage{section}
\begin{document}
\renewcommand\arraystretch{1.1}
\setlength{\abovecaptionskip}{0.1cm}
\setlength{\belowcaptionskip}{0.5cm}
\title { Systematic study of the energy of vanishing flow:
Role of equations of state and cross sections}
\author {Aman D. Sood and  Rajeev K. Puri\\
\it Department of Physics, Panjab University, Chandigarh -160 014,
India.\\}
\maketitle
\begin{abstract}
We present a systematic study of the energy of vanishing flow by
considering symmetric colliding nuclei (between $^{12}$C and
$^{238}$U) at normalized impact parameters using variety of
equations of state (with and without momentum dependent
interactions) as well as different nucleon-nucleon cross sections.
A perfect power law mass dependence is obtained in all the cases
which passes through calculated points nicely. Further, the choice
of impact parameter affects the energy of vanishing flow
drastically, demanding a very accurate measurement of the impact
parameter. However, the energy of vanishing flow is less sensitive
towards the equation of state as well as its momentum dependence.
\end{abstract}
 Electronic address:~rkpuri@pu.ac.in
\newpage
\section{Introduction}
The heavy-ion collisions around Fermi energies have been a subject
of intensive theoretical and experimental investigations in recent
times
\cite{trau04,ogil90,west93,cuss02,mage0062,ange97,buta95,sull90,pak97,krof91,he96,westnpa,krof92,mage0061,zhan90,xu92,li93,zhou94,mota92,lehm96,kuma98,sood04}.
This is primarily due to the onset of multifragmentation
\cite{puri98}, collective flow and its disappearance
\cite{ogil90,west93,cuss02,mage0062,ange97,buta95,sull90,pak97,krof91,he96,westnpa,krof92,mage0061,zhan90,xu92,li93,zhou94,mota92,lehm96,kuma98,sood04}
as well as a mixture of fusion, decay, and fission \cite{puri98}
dominating the physics at these incident energies. Among all these
phenomena, collective flow has been found to be sensitive towards
the nuclear matter equation of state (EOS) as well as towards
nucleon-nucleon (nn) cross section
\cite{ogil90,west93,cuss02,mage0062,ange97,buta95,sull90,pak97,krof91,he96,westnpa,krof92,mage0061,zhan90,xu92,li93,zhou94,mota92,lehm96,kuma98,sood04}.
At low incident energies, dominance of attractive mean field
prompts the emission of particles into backward hemi-sphere
whereas particle emission at higher incident energies is dominated
by the forward scattering. In other words, the net transverse flow
disappears at a certain incident energy termed as the energy of
vanishing flow (EVF)
\cite{ogil90,west93,cuss02,mage0062,ange97,buta95,sull90,pak97,krof91,he96,westnpa,krof92,mage0061,zhan90,xu92,li93,zhou94,mota92,lehm96,kuma98,sood04}.
This energy of vanishing flow has been of rich importance because
of the fact that it is independent of the particles under study
\cite{west93} and is also insensitive towards the apparatus
corrections \cite{west93,cuss02}. This energy of vanishing flow
has also been robusted for the search of equation of state
\cite{mage0062}. Up to now, one has measured the energy of
vanishing flow in the reactions of $^{12}$C+$^{12}$C
\cite{west93}, $^{20}$Ne+$^{27}$Al \cite{west93},
$^{36}$Ar+$^{27}$Al \cite{ange97,buta95}, $^{40}$Ar+$^{27}$Al
\cite{sull90}, $^{40}$Ar+$^{45}$Sc \cite {west93,pak97,mage0062},
$^{40}$Ar+$^{51}$V \cite {krof91}, $^{64}$Zn+$^{27}$Al \cite
{he96}, $^{40}$Ar+$^{58}$Ni \cite {cuss02}, $^{64}$Zn+$^{48}$Ti
\cite {buta95}, $^{58}$Ni+$^{58}$Ni
\cite{cuss02,mage0062,westnpa}, $^{58}$Fe+$^{58}$Fe
\cite{westnpa}, $^{64}$Zn+$^{58}$Ni \cite {buta95},
$^{86}$Kr+$^{93}$Nb \cite {west93,mage0062}, $^{93}$Nb+$^{93}$Nb
\cite {krof92}, $^{129}$Xe+$^{118}$Sn \cite {cuss02},
$^{139}$La+$^{139}$La \cite {krof92}, and $^{197}$Au+$^{197}$Au
\cite {mage0062,mage0061,zhan90}. A power law mass dependence of
the energy of vanishing flow has also been reported in theoretical
\cite{west93,mage0062,westnpa,mage0061,zhou94,sood04} as well as
in experimental studies \cite{west93,mage0062,westnpa,mage0061}.
Looking at these mass dependence analyses, one observes that the
power law behaviour reported for the mass dependence is not
absolute, instead, it explains an average general trend of the
mass dependence
\cite{west93,mage0062,westnpa,mage0061,zhou94,sood04}. For
example, the maximum deviation of the experimental energy of
vanishing flow from the power law is $\approx 18\%$ \cite{sood04}
whereas it varies between 8-19\% in theoretical analysis
\cite{west93,sood04}. These cases are not isolated ones, rather,
they occur very frequently. These huge deviations from the mass
power law can be due to the following reasons.

(i) The choice of impact parameter is not uniform in the above
listed reactions. It varies from $b/b_{max}=0.1$ to 0.4
($b/b_{max}=R_{1}+R_{2}$; $R_{i}$ is the radius of either target
or projectile). It has been reported by many authors that the
energy of vanishing flow depends significantly upon the impact
parameter \cite{mage0062,sull90,pak97,he96}. For example, the
measured energy of vanishing flow varies from $84\pm 7$
MeV/nucleon to $104\pm 5$ MeV/nucleon for the reaction of
$^{40}$Ar+$^{45}$Sc between $b/b_{max}=0.28$ and 0.48
\cite{pak97}. Whereas theoretically, it varies from 65 MeV/nucleon
to 90 MeV/nucleon. Therefore, it is utmost important to have a
concrete and accurate information about the impact parameter.

(ii) The asymmetric reacting partners do not follow the power law
behaviour of symmetric nuclei at the first place. The asymmetry of
the reaction plays a dramatic role in heavy-ion observables.
Therefore, one can not expect to have an absolute power law
behaviour for a mixture of symmetric and asymmetric nuclei as
reported in Refs. \cite{west93,mage0062,mage0061,sood04}. In the
above mass dependent studies, the asymmetry of colliding nuclei
was as high as 0.4 (=$|\frac{A_{1}-A_{2}|}{A_{1}+A_{2}}|$). For a
better understanding, one needs to have a complete systematic
study of the mass dependence of energy of vanishing flow where
impact parameter as well as asymmetry of the reacting nuclei is
kept under control so that the above mentioned ambiguities can be
removed. Our present aim, therefore, is at least twofold.

(a) To present a systematic study of the mass dependence of energy
of vanishing flow by analyzing symmetric reacting partners at
fixed reduced impact parameters. This will remove all the above
mentioned discrepancies.

(b) As reported in the literature, the studies of energy of
vanishing flow give confused picture about the equation of state,
momentum dependent interactions (MDI) as well as about the
in-medium nn cross section. Some studies suggest hard equation of
state producing more flow than soft
\cite{west93,mage0062,he96,zhou94,lehm96} whereas others have
contradictory experiences
\cite{ogil90,west93,mage0062,westnpa,krof92}. Some studies deny
the role of momentum dependent interactions \cite{west93,mage0062}
whereas others point toward their dominant role
\cite{pak97,westnpa,zhou94}. The above systematic study shall also
be extended to shed light on the equation of state, momentum
dependent interactions as well as on the in-medium nn cross
section throughout the periodic table, for the first time. This
study will also help to clear many unresolved questions about the
nuclear matter equation of state as well as in-medium nn cross
section. The above study is carried out
within the framework of quantum molecular dynamics (QMD) model \cite{aich91}.\\
\section{The model}
In the QMD model \cite{aich91}, each nucleon propagates under the
influence of mutual two and three-body interactions. The
propagation is governed by the classical equations of motion:
\begin{equation}
\dot{{\bf r}}_i~=~\frac{\partial H}{\partial{\bf p}_i}; ~\dot{{\bf
p}}_i~=~-\frac{\partial H}{\partial{\bf r}_i},
\end{equation}
where H stands for the Hamiltonian which is given by:
\begin{equation}
H = \sum_i^{A} {\frac{{\bf p}_i^2}{2m_i}} + \sum_i^{A}
({V_i^{Skyrme} + V_i^{Yuk} + V_i^{Coul} + V_i^{mdi}}).
\end{equation}
Here $V_{i}^{Skyrme}$, $V_{i}^{Yuk}$, $V_{i}^{Coul}$, and
$V_i^{mdi}$ are, respectively, the Skyrme, Yukawa, Coulomb, and
momentum dependent potentials. The momentum dependent interactions
are obtained by parameterizing the momentum dependence of the real
part of the optical potential. The final form of the potential
reads as \cite{aich91}
\begin{equation}
U^{mdi}\approx t_{4}ln^{2}[t_{5}({\bf p_{1}}-{\bf
p_{2}})^{2}+1]\delta({\bf r_{1}}-{\bf r_{2}}).
\end{equation}
Here $t_{4}$ = 1.57 MeV and $t_{5}$ = $5\times 10^{-4} MeV^{-2}$.
A parameterized form of the local plus momentum dependent
interaction potential is given by
\begin{equation}
U=\alpha \left({\frac {\rho}{\rho_{0}}}\right) + \beta
\left({\frac {\rho}{\rho_{0}}}\right)^{\gamma}+ \delta
ln^{2}[\epsilon(\rho/\rho_{0})^{2/3}+1]\rho/\rho_{0}.
\end{equation}

The parameters $\alpha$, $\beta$, $\gamma$, $\delta$, and
$\epsilon$ are listed in Ref. \cite{aich91}. \\
\section{Results and discussion}
The present systematic study is carried out by simulating the
symmetric reactions of $^{12}$C+$^{12}$C, $^{20}$Ne+$^{20}$Ne,
$^{40}$Ca+$^{40}$Ca, $^{58}$Ni+$^{58}$Ni, $^{93}$Nb+$^{93}$Nb,
$^{131}$Xe+$^{131}$Xe, $^{197}$Au+$^{197}$Au, and
$^{238}$U+$^{238}$U at their corresponding reduced impact
parameters $b/b_{max}=0.4$ and 0.2 using the hard, soft, and hard
equation of state with momentum dependent interactions. We shall
also use constant nn cross sections of 40 and 55 mb strength for
the present analysis. The static hard and soft equations of state
are labeled as Hard and Soft whereas momentum dependent inclusion
is written as HMD and SMD, respectively. The nn cross sections are
written as superscripts to these abbreviations. The above
mentioned reactions were simulated at incident energies between 30
MeV/nucleon and 250 MeV/nucleon in small steps and energy of
vanishing flow was calculated using straight line interpolations.
In Table 1, we list the incident energies and corresponding
transverse flow over which straight line interpolation was made to
find out the energy of
vanishing flow.\\

\begin{figure}[!t]
\centering
 \vskip -1cm
\includegraphics[width=13cm]{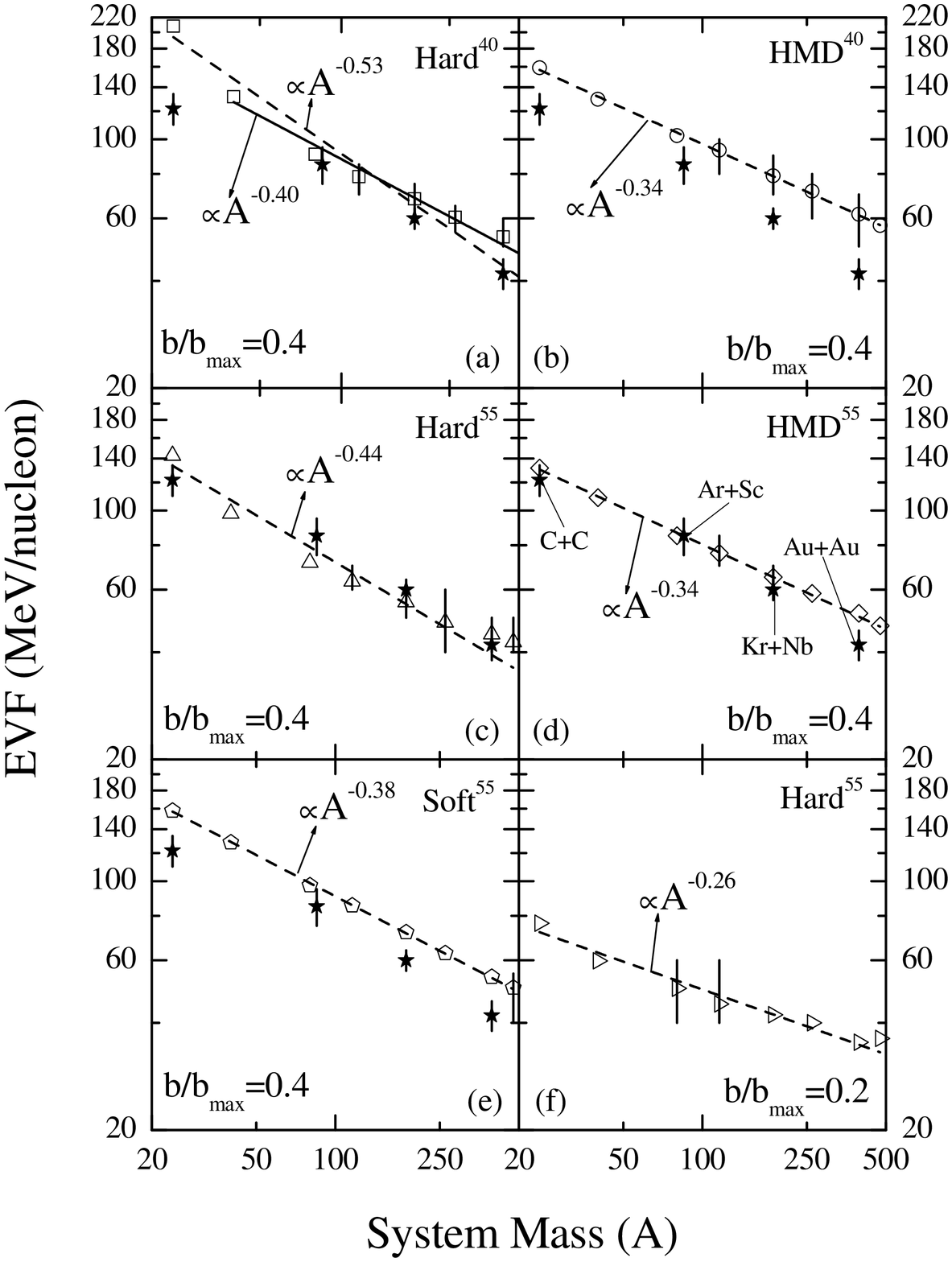}
\caption{The energy of vanishing flow as a function of combined
mass of the system for different model ingredients. Parts
\emph{a}, \emph{b}, \emph{c}, \emph{d}, and \emph{e} are for
$b/b_{max}=0.4$ whereas part \emph{f} is for $b/b_{max}=0.2$. Open
symbols represent our theoretical calculations whereas solid stars
represent the experimental data at $b/b_{max}=0.4$. Lines are
power law fit $\propto A^{\tau}$. The solid line in part \emph{a}
is a power law fit excluding $^{12}$C+$^{12}$C
reaction.}\label{fig1}
\end{figure}

In Fig. \ref{fig1}, we display the energy of vanishing flow as a
function of combined mass of the system (between 24 and 476) for
different equations of state and nn cross sections along with
available experimental data.
First of all, baring few cases, all the calculated energies of
vanishing flow fall on the same line that is a fit of the power
law nature $\propto A^{\tau}$. This is very encouraging since as
has been reported in Refs. \cite{west93,mage0062,mage0061,sood04},
these power law dependencies were not absolute and actual energy
of vanishing flow was as far as by 23 MeV/nucleon from the mass
power law fit. Here we notice that the power law curve passes
through all the calculated points justifying our above discussion
that the deviation from the power law behaviour was due to
different impact parameters and asymmetric colliding nuclei. While
comparing Figs. \ref{fig1}c and \ref{fig1}f, one also sees the
role of impact parameter in energy of vanishing flow. We see a
significant variation in the value of energy of vanishing flow
while going from $b/b_{max}=0.4$ to 0.2. This is particularly true
for lighter colliding nuclei. For example, one sees that for the
reaction of $^{12}$C+$^{12}$C, the energy of vanishing flow at
$b/b_{max}=0.4$ is 143 MeV/nucleon which reduces to 76 MeV/nucleon
for $b/b_{max}=0.2$. This huge variation in the energy of
vanishing flow due to impact parameter variation was not glorified
in earlier publications of mass dependence
\cite{west93,mage0062,mage0061,zhou94,sood04}. This huge variation
also explains why mass dependence of energy of vanishing flow is
not explained by power laws reported in earlier publications.
\begin{table}
\caption{The incident energy (MeV/nucleon) as well as collective
transverse flow (MeV/\emph{c}) for all the reactions reported
above using $Hard^{55}$ at $b/b_{max}=0.4$ and 0.2 and $HMD^{55}$
at $b/b_{max}=0.4$.}
\begin{center}
\begin{tabular}{|c|c|c|c|c|c|c|} \hline
&\multicolumn{4}{c|}{$b/b_{max}=0.4$}&\multicolumn{2}{c|}{$b/b_{max}=0.2$}\\\cline{2-7}
System&\multicolumn{2}{c|}{$Hard^{55}$}&\multicolumn{2}{c|}{$HMD^{55}$}&\multicolumn{2}{c|}{$Hard^{55}$}\\\cline{2-7}
            &Energy  &$<p_{x}^{dir}>$  &Energy &$<p_{x}^{dir}>$&Energy  &$<p_{x}^{dir}>$\\\cline{1-7}
        $^{12}$C+$^{12}$C    &140  &-0.43 &130 &-0.53 &70  &-1.57  \\\cline{2-7}
                             &160  &2.57  &140 &2.52  &80  &0.98   \\
 \hline $^{20}$Ne+$^{20}$Ne  &90   &-1.35 &100 &-2.81 &50  &-3.25  \\\cline{2-7}
                             &100  &0.32  &110 &0.42  &60  &0.1    \\
 \hline $^{40}$Ca+$^{40}$Ca  &65   &-2.78 &80  &-2.25 &40  &-3.32  \\\cline{2-7}
                             &75   &1.53  &90  &2.38  &60  &3.37   \\
 \hline $^{58}$Ni+$^{58}$Ni  &60   &-1.37 &70  &-2.27 &40  &-1.48  \\\cline{2-7}
                             &70   &2.8   &85  &3.54  &60  &4.22   \\
 \hline $^{93}$Nb+$^{93}$Nb  &50   &-2.34 &60  &-2.19 &40  &-0.61  \\\cline{2-7}
                             &60   &2.05  &70  &2.13  &50  &2.23   \\
 \hline $^{131}$Xe+$^{131}$Xe&40   &-3.41 &55  &-1.47 &40  &-0.01  \\\cline{2-7}
                             &60   &4.6   &65  &2.75  &60  &8.03   \\
 \hline $^{197}$Au+$^{197}$Au&40   &-2.22 &50  &-0.68 &35  &-0.09  \\\cline{2-7}
                             &50   &2.3   &55  &1.6   &50  &5.06   \\
 \hline $^{238}$U+$^{238}$U  &40   &-1.06 &45  &-1.07 &35  &-0.4   \\\cline{2-7}
                             &50   &2.82  &55  &3.26  &40  &1.3    \\
 \hline
\end{tabular}
\end{center}
\end{table}
This also points toward the need of accurate impact parameter
analysis in experimental measurements where one often either gives
the range or the upper value of the estimated impact parameters.
The insufficient and inaccurate estimates of impact parameters in
experiments can play a havoc for pinning down the nuclear matter
equation of state from the study of energy of vanishing flow. For
instance, Hard$^{55}$ and Soft$^{55}$ equations of state at
$b/b_{max}=0.4$, for $^{12}$C+$^{12}$C reaction give energy of
vanishing flow equal to 143 and 158 MeV/nucleon whereas
Hard$^{55}$ at $b/b_{max}=0.2$ has an energy of vanishing flow
equal to 76 MeV/nucleon. A small error in the estimation of impact
parameter can lead to completely different interpretation of the
nature of equation of state and other physical conclusions.\\
From the figure, it is also evident that the average difference
between Hard and HMD equations of state (for $\sigma = 40$ and 55
mb) is 15\% whereas for hard and soft equations of state, it is
26\%. On the other hand, it is 42\% for hard equation of state at
$b/b_{max}=0.4$ and 0.2. The role of momentum dependent
interactions is stronger for lighter colliding nuclei compared to
heavier nuclei due to larger energy of vanishing flow in lighter
cases in agreement with Ref. \cite{zhou94}. Further, larger cross
section leads to smaller energy of vanishing flow which is in
agreement with Refs. \cite{west93,he96,xu92,sood04}. The average
role of momentum dependent interactions (in deciding the energy of
vanishing flow) is smaller due to low energy of vanishing flow.
This is also in agreement with Ref. \cite{pak97}.  This may,
however be, significant for peripheral reactions where energy of
vanishing flow with static equation of state is much higher than
the experimental values \cite{pak97}. One also sees that both the
Hard$^{55}$ and HMD$^{55}$ equations of state are able to explain
the energy of vanishing flow (including $^{12}$C+$^{12}$C)
throughout the periodic table at $b/b_{max}=0.4$ whereas this is
not possible with smaller cross section. It is worth mentioning
that the $^{12}$C+$^{12}$C reaction is very rarely discussed in
the literature. One also sees that the power factors $\tau$ in all
the cases are close to each other expect for Hard$^{40}$. In all
the other cases, these are close to -1/3 defending the mutual
dominance of attractive nuclear forces $\propto 2/3$ and repulsive
nuclear forces $\propto$ unity, therefore, leading to net -1/3
dependence. We have also tested the role of asymmetry of colliding
nuclei on energy of vanishing flow which is also found to be
significant.
\begin{figure}[!t] \centering
 \vskip -1cm
\includegraphics[width=10cm]{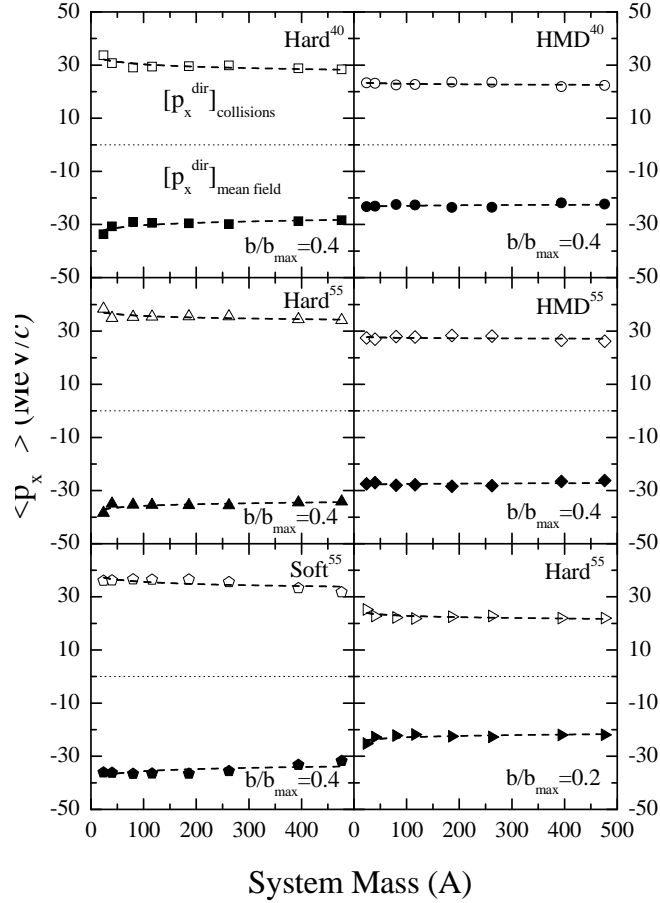}
\caption{The decomposition of collective transverse in-plane flow
at energy of vanishing flow into mean field (solid symbols) and
collision parts (open symbols). Lines are power law fit $\propto
A^{\tau}$}\label{fig2}
\end{figure}

In Fig. \ref{fig2}, we display the division of collective in-plane
flow at energy of vanishing flow into parts resulting from the
repulsive nn collisions and attractive mean field. We see that the
mean field is always attractive whereas it is counterbalanced by
the repulsive collision part leading to net zero collective
transverse in-plane flow at the energy of vanishing flow. This
figure also depicts the mass independent nature of the
contribution of mean field towards collective transverse in-plane
flow at the energy of vanishing flow. This is in agreement with
earlier calculation \cite{sood04}.
\section{Summary}
 Summarizing, we presented the systematic study
of the energy of vanishing flow by considering symmetric colliding
nuclei at normalized impact parameters using variety of equations
of state (with and without momentum dependent interactions) as
well as different cross sections. We obtained a perfect power law
mass dependence in all the cases which passes through all the
calculated points nicely throughout the periodic table. Further,
the choice of impact parameter affects the energy of vanishing
flow drastically, demanding a very accurate measurement of impact
parameter. However, the energy of vanishing flow is found to be
less sensitive towards the equation of state as well as its
momentum dependence.

\end{document}